\documentclass[%
 twocolumn,
 superscriptaddress,
 amsmath,amssymb,
 aps,
]{revtex4-2}

\usepackage{tikz}
\usepackage{caption}
\usepackage{appendix}
\usepackage{hyperref}

\usetikzlibrary{patterns, hobby, matrix}
\usepackage{pgfplots}

\pgfplotsset{compat=1.6}

\begin{document}

\preprint{APS/123-QED}

\title{Non-Newtonian rheology in a capillary tube with varying radius }

\author{Federico Lanza}
\author{Alberto Rosso}
\affiliation{LPTMS, CNRS, Université Paris-Saclay, 91405 Orsay, France}%
\author{Laurent Talon}
\affiliation{FAST, CNRS, Université Paris-Saclay, 91405, Orsay, France}%
\author{Alex Hansen}
\affiliation{PoreLab, Department of Physics, Norwegian University of Science and Technology, N-7491 Trondheim, Norway}%


\begin{abstract}
The flow through a capillary tube with non-constant radius and where bubbles of yield stress fluid are injected is strongly non-linear.
In particular below a finite yield pressure drop, $P_y$, flow is absent, while a singular behaviour is expected above it. In this paper we compute the yield pressure drop statistics and the mean flow rate in two cases: (i) when a single bubble is injected, (ii) when many bubbles are randomly injected in the fluid.
\end{abstract}


\maketitle

\section*{Introduction}
In many industrial, geophysical or biological applications related to porous media, non-Newtonian fluids are frequently encountered. Indeed many complex fluids present a non-linear rheology as for example slurries, heavy oils, suspensions \cite{barnes89,coussot05} or some biological fluids like blood  \cite{popel05,bessonov16}. Here, we are interested in yield stress fluids, which require a minimal applied stress to flow. These fluids are involved in many practical applications, such as drilling for oil extraction, where proppant fluids are injected in the soil for the fracking process \cite{Barbati2016}, stabilization of bone fractures in biomedical engineering \cite{WidmerSoyka2013}, or ground reinforcement by cement injection. Yield stress fluids in porous media is a challenging and interesting problem which has been the subject of many studies in the last decades \cite{entov67,park73,al-fariss87,chen05,sochi08,talon13b,rodriguezdecastro17,liu19}. Because of the presence of a yield stress, the fluid is able to flow only if a certain amount of pressure is imposed \cite{roux87,chen05,liu19,fraggedakis21}. There is then a strong coupling between the rheology of the fluid and the disorder of the porous structure, implying that some regions are easier to yield than others. Above this pressure threshold, as demonstrated by several studies \cite{roux87,talon13b,chevalier15a, waisbord19,liu19}, a progressive increase of flowing paths occurs. As a consequence, the flow rate increases with the applied pressure according to a power-law:
\begin{equation}
 Q \propto (\Delta P -  P_y)^\beta.
 \label{eq:PL}
\end{equation}
The origin of this flowing regime is an effect of the disorder, but remarkably the value of the exponent $\beta=2$ is independent of the type of disorder in 2D porous media \cite{liu19}. 

This is however not the case in 1D, if one describes the porous media by a series of uniform bundle of capillaries. The flow curve above the threshold depends then on the details of the opening distribution \cite{nash16}.

If the flow of yield stress fluids in porous media is already a challenging problem, in many situations the complexity is increased by the presence of different immiscible fluids.
Multi-phase flow in porous media is a very old and rich subject, and is still the topic of many ongoing research. One of the main difficulties lies in the presence of numerous interfaces exerting capillary forces on the fluids present, which makes the dynamic very non-linear.
It is then surprising that, during many decades, the models predicting the mean flow rate as function of the mean applied pressure had assumed linear relations \cite{bear88,dullien91}.

In the last decade, however, a series of experiments and simulations \cite{tallakstad09,rassi11,sinha12,yiotis13,chevalier15b,zhang21} have shown the existence of a non-linear flowing regime at low flow rate. Similarly to the yield stress fluid case, the physical reason behind this observation lies in the presence of the heterogeneity. In fact, due to capillary forces the interfaces can only move if a certain pressure is applied. In a disordered media, certain regions allows the movement of interfaces more easily than others. At very low applied pressure, the displacement of the interfaces occurs only in few pathways whose number raises with the applied pressure \cite{yiotis19}. This increase is then responsible for a non-linear flow rate-pressure relationship similar to eq. \eqref{eq:PL}, where the exponent $\beta$ has been reported to vary in the range $\beta\in[1.5,2]$ depending on the flow condition \cite{tallakstad09,yiotis13,rassi11,sinha12,sinha17,yiotis19,zhang21}.  An argument based on
comparing length scales associated with the viscous forces compared with those set by the capillary forces gave $\beta=2$ \cite{tallakstad09}. This value was also found using a mean field theory-based calculation \cite{sinha12}, whereas a calculation based on the capillary fiber
bundle model, gives either $\beta=3/2$ or $\beta=2$ depending on the statistical distribution of the flow thresholds \cite{Roy2019}. These approaches are all based on the mobilization of interfaces and can be understood for instance by considering many bubbles in a single 1D pore with spatial varying opening. In this case, there exists a minimal pressure threshold to initiate the flow. Above, the flow rate increases with a power-law as eq. \eqref{eq:PL}, similarly to a Herschel-Bulkley yield stress fluid with index $n=1/\beta$. 

Ausrsj{\o} et al. \cite{aetfhm14} find an exponent $\beta = 1.49$ and 1.35 in a two-dimensional model porous medium where transport of one of the fluids occurs entirely through film flow, depending on the fractional flow rate.     

In this work, we aim to investigate  two-phase flows, but in the case where one of the two fluids presents a yield threshold. The situation is then more complex as both the rheology and the surface tension lead to a threshold pressure to initiate the flow. For simplification, we propose to model a porous medium by a set of identical capillaries with varying opening (e.g. fibre bundle model).
The first question we want to address is the determination of the pressure threshold depending on the rheology, the surface tension and the disorder of the capillaries.
The second question is then to determine the flow curve just above this threshold. We will show that the flow rate follows a power law, and we will determine the exponent depending the structure disorder.\\

The volumetric flow rate, $q$, is expected to grow linearly with the pressure gradient $\Delta P/ l$. Here $\Delta P = P_{\rm{in}} - P_{\rm{out}}$ is the pressure difference applied to the edges of a tube of length $l$ (in the following we assume $\Delta P>0$ for simplicity). This behaviour is recovered for Newtonian fluids in a cylindrical capillary tube of radius $r_0$ and given by the celebrated Poiseuille law $q=\pi r_0^4 \Delta P/ (8 \mu l)$, where $\mu$ is the fluid viscosity.
However, non-Newtonian yield stress fluids display a non-linear response. Their rheology can be modeled by the Herschel–Bulkley constitutive equation \cite{bird76}, that gives a relation between the shear stress $\tau$ applied to the fluid and the shear rate $\dot{\gamma}$
\begin{equation}
    \tau = \tau_c + k\dot{\gamma}^n,
    \label{eq:HB_model}
\end{equation}
The constant $k$ is the consistency, the exponent $n>0$ is the flow index and $\tau_c$ is the yield stress. In this case the flow in the tube occurs only above an yield pressure drop $P_y$, and the flow rate grows with pressure in a non-linear way. For example, for a perfect cylindrical tube filled with a non-Newtonian yield stress fluid, the yield pressure is $ P_{y} = 2\tau_c l/r_0$ and the flow law \cite{bird87}:
\begin{equation}
    q  =
    \begin{cases}
    C_0\, r_0^{4+\frac{1}{n}}\left( \frac{\Delta P -  P_{y}}{l} \right)^{\frac{1}{n} + 1} \hspace{0.2cm} \text{if}\hspace{0.2cm} \Delta P \rightarrow P_y^+,\\
    \\
    C_{\infty}\, r_0^{3+\frac{1}{n}} \left( \frac{\Delta P -   \widetilde P_y}{l} \right)^{\frac{1}{n}}  \hspace{0.4cm} \text{if}\hspace{0.2cm} \Delta P\gg  P_{y},\\
    \end{cases}    
    \label{eq:q_nonnewtonian_cyl}
\end{equation}
where $C_0 =n \pi /( (n+1)2^{1+1/n} k^{1/n}\tau_c)$, $C_{\infty} = n \pi / ((3n +1)(2k)^{1/n})$ and $ \widetilde P_y = ((3n+1)/(2n+1)) P_y$ a pseudo critical pressure (see \cite{talon14,Bauer19}).\\

\section{Model for a single bubble}

We now consider the case of a tube filled with a Newtonian liquid in which one small bubble of yield stress fluid (YSF) is injected. We assume the fluids to be immiscible and incompressible.
The bubble, of size $\Delta x_b \ll l$ and position $x_b$, is at the origin of a critical yield pressure $P_{y} = 2\tau_c \Delta x_b/r_0$. The total pressure drop $\Delta P$ needed to sustain a flow rate $q$ can be expressed as the sum of the pressure drops across every portion of fluid. The pressure drops across both portions of Newtonian fluid, in the intervals $0 < x < x_b$ and $x_b+\Delta x_b< x < l$, are given by the Poiseuille law
\begin{equation}
    \begin{split}
    &P_{\rm{in}} - P_{x_b}^- = q \frac{8\mu x_b}{\pi r_0^4},\\
    &P_{x_b + \Delta x_b}^+ - P_{\rm{out}} = q\, \frac{8\mu (l - x_b - \Delta x_b)}{\pi r_0^4}.
    \end{split}
    \label{eq:P_Newtonian_cyl}
\end{equation}
The pressure drop across the bubble is instead given by Equation (\ref{eq:q_nonnewtonian_cyl}) and writes
\begin{equation}
    P_{x_b}^+ - P_{x_b + \Delta x_b}^{-} =
    \begin{cases} \left(\frac{q}{C_0\, r_0^{4+\frac{1}{n}}}\right)^{\frac{n}{n+1}}\! \Delta x_b + P_{y}  \hspace{0.2cm} \text{if}\hspace{0.2cm} \Delta P \rightarrow P_y^+,\\
    \\
    \hspace{0.7cm} q^n \frac{\Delta x_b}{C_{\infty}^n r_0^{3n+1}}+ \tilde P_{y} \hspace{0.75cm} \text{if}\hspace{0.2cm} \Delta P\gg P_{y}\\
    \end{cases}
    \label{eq:P_bubble_cyl}
\end{equation}
Moreover, when two immiscible fluids are in contact, at the interface emerges a discontinuity in pressure, called capillary pressure, whose sign depends on the curvature of the interface \cite{bear88}. Hence,
in a perfect cylindrical tube, the total capillary pressure across the two interfaces of a bubble cancels out since at each interface the capillary pressure discontinuity is $2\sigma/r_0$ ($\sigma$ being the surface tension between the two fluids), but the signs of the two contributions are opposite as the two interfaces have opposite curvature \cite{note1}.\\
The sum of the three pressure drops given in equations \eqref{eq:P_Newtonian_cyl} and \eqref{eq:P_bubble_cyl} in the limit $P \gtrsim P_{y}$ is then
\begin{equation}
    \Delta P  = q \frac{8\mu  (l - \Delta x_b)}{\pi r_0^4}+q^{\frac{n}{n+1}} \frac{\Delta x_b}{C_0^{\frac{n}{n+1}} r_0^{\frac{4n+1}{n+1}}} + P_{y}.
    \label{eq:P_near_cyl}
\end{equation}
In this limit the flow vanishes to 0, so we can neglect the linear term in equation \eqref{eq:P_near_cyl} as $n/(n+1) < 1\ \forall \ n > 0$.\\
In the opposite limit $P\gg P_y$ we have
\begin{equation}
    \Delta P  = q \frac{8\mu  (l - \Delta x_b)}{\pi r_0^4} + q^{n} \frac{\Delta x_b}{ C_{\infty}^n r_0^{3n+1} } + \tilde P_{y}.
    \label{eq:P_far_cyl}
\end{equation}
Since now $q\to \infty$, we should distinguish between a shear-thinning fluid and a shear-thickening fluid, for which $n<1$ and $n>1$ respectively. In the first case, the leading term is the one proportional to $q^n$, while in the other case the leading term is the linear one. Finally, we can write the volumetric flow rate in the two different limits:
\begin{equation}
    q(\Delta P) = \begin{cases} 
    \hspace{0.8cm}C_0\,r_0^{4 + \frac{1}{n}} \left(\frac{\Delta P - P_{y}}{\Delta x_b}\right)^{1+\frac{1}{n}} \hspace{0.8cm} \text{if} \hspace{0.2cm} q\to 0\\
    \\
    \begin{cases}
    C_{\infty} r_0^{3+\frac{1}{n}} \left(\frac{\Delta P - \widetilde P_y}{\Delta x_b}\right)^{\frac{1}{n}}\hspace{0.1cm} \text{if} \hspace{0.2cm} n<1\\
    \\
    \hspace{0.7cm}\frac{\pi r_0^4}{8\mu} \frac{\Delta P - \widetilde P_{y}}{l - \Delta x_b}\hspace{0.75cm} \text{if} \hspace{0.2cm} n>1\\
    \end{cases} \text{if} \hspace{0.2cm} q\to +\infty\\
    \end{cases}
    \label{eq:q_cyl}
\end{equation}

\subsection{Non-uniform tube}
\begin{figure}
    \centering
    \includegraphics[scale=0.5]{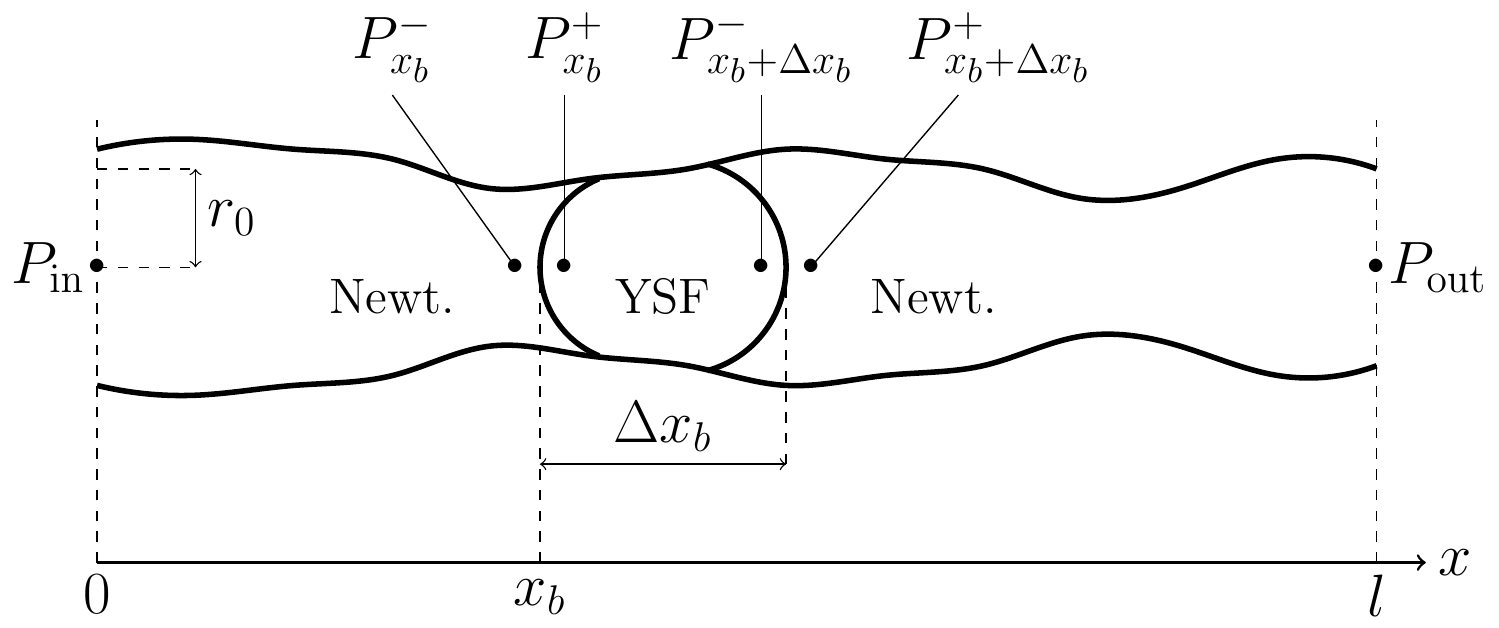}
    \caption{Sketch of a bubble of yield stress fluid in a non-uniform tube of length $l$.}
    \label{fig:sketch_1bubble}
\end{figure}

We consider now a tube still of length $l$, but with varying radius $r(x)$ (see figure \ref{fig:sketch_1bubble}) described by the following equation 
\begin{equation}
    r(x) = \frac{r_0}{1+af(x)},
    \label{eq:r_general}
\end{equation}
where $f(x)$ is a bounded function with zero average in the interval $x\in[0,l]$, $a \ll 1$ a dimensionless constant and $r_0$ a characteristic radius.
The radius varies slowly enough so that the radial component of the fluid velocity can be neglected with respect to the axial one (lubrication limit, see \cite{Frigaard04}), and the bubble size $\Delta x_b$ can be considered constant along the tube, as its modification are of the order of $a$. Nevertheless, the spatial variation of the tube geometry affects both the flow curve and the value of the yield pressure.\\
In particular, two modifications should be included. First, the capillary pressure across the bubble interfaces do not cancel anymore \cite{Sinha2013}. Since $P_{x_b}^- - P_{x_b}^+ = 2 \sigma/r(x_b)$ and $P_{x_b + \Delta x_b}^{+} - P_{x_b + \Delta x_b}^{-} = 2 \sigma/r(x_b + \Delta x_b)$, their difference is in general non zero and approximately equal to
\begin{equation}
  \frac{2\sigma}{r(x_b)} - \frac{2\sigma}{r(x_b+\Delta x_b)} \simeq a P_\sigma  \Delta x_b f'(x_b),  
    \label{eq:Pt_general}
\end{equation}
with $P_\sigma = 2 \sigma/r_0$.
Secondly, as $r(x)$ is non-constant, both Poiseuille law and Eq. (\ref{eq:q_nonnewtonian_cyl}) can be considered valid only along infinitesimal intervals of length $dx$. For both reasons, the flow rate varies in time as a function of the bubble location $x_b$. The Poiseuille equation becomes $q(x_b) =-\pi r(x)^4 (dP/dx)/(8\mu)$ from which, at the first order of $a$, we get:
\begin{eqnarray}
    &&P_{\rm{in}} - P_{x_b}^- = q(x_b)\,\frac{8\mu}{\pi r_0^4} \left(x_b + 4a\!\int_0^{x_b}\!f(x)dx \right),\\
    &&P_{x_b + \Delta x_b}^+ - P_{\rm{out}}  =\nonumber\\
    &&= q(x_b)\,\frac{8\mu}{\pi r_0^4}\! \left(l - x_b - \Delta x_b + 4a\!\int_{x_b+\Delta x_b}^{l}\!f(x)dx\right) 
    \label{eq:P_Newtonian}
\end{eqnarray}
Using the differential form of eq. \eqref{eq:q_nonnewtonian_cyl} in the limit of small flow rate, we obtain:
\begin{equation*}
    \begin{split}
    P_{x_b} - P_{x_b + \Delta x_b} &\simeq \left(\frac{q(x_b)}{C_0\, r_0^{4+\frac{1}{n}}}\right)^{\frac{n}{n+1}} \Delta x_b +
    P_y^0 \left(1 + a f(x_b)\right),
    \end{split}
    \label{eq:P_bubble}
\end{equation*}
where $P_y^0= 2 \tau_c \Delta x_b / r_0$. Note that we approximated the integral:
\begin{equation}
\int_{x_b}^{x_b+\Delta x_b} \frac{1}{r^{4+1/n}(x)}dx \simeq \frac{1}{r_0^{4+1/n}} \Delta x_b,
\end{equation}
because the correction only affects the prefactor of the flow curve, and not the exponent or the threshold.
Since in this limit $q(x_b) \ll 1$, the leading behavior of the flow curve can be written as:
\begin{equation}
    q(x_b) \simeq C_0\, r_0^{4 + \frac{1}{n}} \left[\frac{\Delta P - \gamma(x_b) }{\Delta x_b} \right]^{1+\frac{1}{n}},
    \label{eq:q_general}
\end{equation}
where $\gamma(x_b) = P^0_{y} + a (P^0_{y} f(x_b) +  P_{\sigma} \Delta x_b f'(x_b))$. From eq. \eqref{eq:q_general} we note that in a deformed tube the critical pressure drop $P_y$ above which flow is possible has increased with respect to the cylindrical tube, and is equal to the maximum of $\gamma(x_b)$:
\begin{equation}
     P_y = P_y^0 +a\max_{0<x_b<l}\left[ P_y^0 f(x_b) + P_{\sigma} \Delta x_b f'(x_b)\right];
    \label{eq:Py_general}
\end{equation}
we denote $x_m$ the position of such maximum. The bubble position moves as $dx_b/dt = q/(\pi r_0^2)$, hence from eq. (\ref{eq:q_general}) we get the equation of motion
\begin{equation}
    \frac{dx_b}{dt} = \frac{C_0\, r_0^{2+\frac{1}{n}}}{\pi \Delta x_b^{1 + \frac{1}{n}}} \left[\Delta P - \gamma(x_b) \right]^{1+\frac{1}{n}}.
    \label{eq:dxdt_general}
\end{equation}
The time $T$ needed for the bubble to move from one end of the tube to the other can be computed from (\ref{eq:dxdt_general}):
\begin{equation}
    T = \int_0^l\! \frac{dx_b}{dx_b/dt} \propto \int_0^l\! \frac{dx_b}{\left[\Delta P - \gamma(x_b) \right]^{1+\frac{1}{n}}}.
    \label{eq:T_general}
\end{equation}
In general $\gamma(x_b)$ relies on the specific form of $f(x_b)$. However, supposing that $f(x_b)$ is analytical, we can expand $\gamma(x_b)$ around $x_m$: $\gamma(x_b) = P_y + \alpha (x_b-x_m)^2 + \dots$\\
For $ \Delta P \to P_y^+$, the dominant contribution to the integral of eq. (\ref{eq:T_general}) is around $x_m$, so we can write
\begin{equation}
    T \propto \int_0^l\! \frac{dx_b}{[ \Delta P - P_y + \alpha (x_b-x_m)^2]^{1+\frac{1}{n}}} \propto (\Delta P-P_y)^{-( \frac{1}{n} + \frac{1}{2} )}.
    \label{eq:T_Pnear}
\end{equation}
The flux averaged over the time $T$, $\langle q \rangle_T$ is then
\begin{equation}
    \langle q \rangle_T = \frac{\pi r_0^2 l}{T} \propto ( \Delta P-P_y)^{\frac{1}{n}+\frac{1}{2}}.
    \label{eq:qmeanT_Pnear}
\end{equation}
Note that close to the yield threshold $P_y$, the power-law exponent $1/n+1/2$ of the flow rate turns out to be different from $1+1/n$ in eq. \eqref{eq:q_cyl} for the uniform tube.\\
On the other hand, in the opposite limit $\Delta P\gg P_y$, since the fluctuations along the critical pressure are negligible, we expect the same behaviour of the cylindrical tube.

As a final remark, we discuss the case where $f(x_b)$ is not analytical. The non-linear prediction of eq. \eqref{eq:qmeanT_Pnear} hold only if $\gamma(x_b)$ is derivable at least twice. Otherwise, its expansion around $x_m$ is of the form:
$\gamma(x_b) = P_y + \alpha |x_b-x_m|^\delta + \dots$, with $\delta > 0$.
In this case, the behavior of the integral in eq. \eqref{eq:T_Pnear} is modified and the flux averaged over $T$ is then
\begin{equation}
    \langle q \rangle_T  \propto ( \Delta P-P_y)^{\frac{1}{n}+1-\frac{1}{\delta}}.
    \label{eq:qmeanT_general}
\end{equation}
To provide a concrete example, we consider a saw-tooth triangular geometry:
\begin{equation}
    f(x) = \frac{4}{l}\left| x -  \frac{l}{2}\right| - 1, \hspace{0.3cm} x\in\left[0,l\right[.
    \label{eq:f_triangle}
\end{equation}
In this case we have
\begin{equation*}
   \gamma(x_b) = P^0_y + \frac{4a}{l} \left[ P^0_y \left| x_b - \frac{l}{2}\right| +  P_{\sigma} \Delta x_b\,\text{sgn}\left(x_b-\frac{l}{2}\right) \right].
   \label{eq:gamma_triangle}
\end{equation*}
Its maximum is located at the discontinuity point $x_m=0$ and writes
\begin{equation}
   P_y = P^0_y + a \left[ 2P^0_y +  \frac{4}{l}P_{\sigma} \Delta x_b \right].
   \label{eq:Py_triangle}
\end{equation}
Integrating equation $\ref{eq:T_general}$ yields to $\delta=1$ if the bubble fluid presents yield stress, while $\langle q \rangle_T  \propto ( \Delta P-P_y)^{\frac{1}{n}+1}$ if the bubbles are Newtonian ($P_y^0 = 0$).

\section{Model for many bubbles}
\begin{figure}
    \centering
    \includegraphics[scale=0.5]{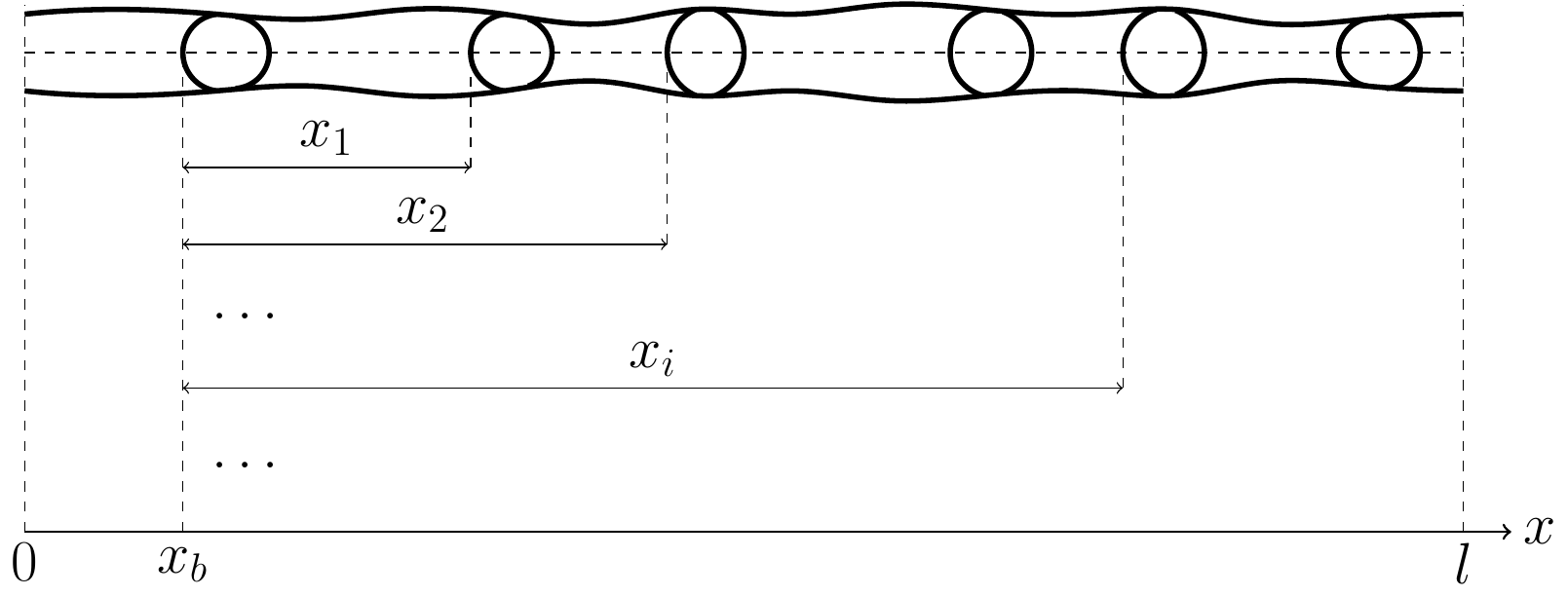}
    \caption{Sketch of several bubbles of yield stress fluid in a non-uniform tube of length $l$.}
    \label{fig:sketch_Nbubbles}
\end{figure}

In a uniform tube, the flow curve obtained when a single shot of non-Newtonian fluid is injected is identical to the one obtained when the same amount of fluid is split in $N$ small bubbles. This is not the case for a non-uniform tube. To be concrete, we address the case of several identical bubbles of non-Newtonian fluid (see figure \ref{fig:sketch_Nbubbles}).
It comes out that the critical pressure obtained with $N$ bubbles of length $\Delta x_b$ is larger than $N P_y^0= 2 N \Delta x_b \,  \tau_c /r_0 $, the value expected for a single shot of length equal to $N \Delta x_b$. The difference depends on the total number of bubbles and on the specific bubble configuration.\\
During the flow the inter-bubble distances remain constant as the fluids are incompressible. Moreover, periodic boundary conditions are set, namely $f(x)=f(x+l)$. This assumption can describe two different situations: (i) a tube of length $l$ with periodic boundary conditions (ii) a tube of length $L\gg l$ presenting a periodic deformation of spatial period $l$. In the latter case, the bubbles are in general located on different periods, but it is convenient to shift their position in the first period: more precisely, if a bubble is located at a certain position in the $k$-th period, the dynamics of the system does not change if we subtract the quantity $(k-1)l$ from that position. We then denote with $x_b$ the position of the most left bubble and with $x_i$ the distance from its $i$-th bubble neighbour. Thus $i=1,\ldots, N-1$, and the $i$-th right neighbour is located at $x_b + x_i$. When $x_b$ moves from $0$ to $l$ all the other bubbles move exactly one period.\\
In the limit of small flow rate $q \to 0$, the pressure drop at the edges of the $i$-th bubble is
\begin{equation}
\begin{split}
    &P_{x_b+x_i}^+ - P_{x_b + x_i + \Delta x_b}^- =\\
    &=  \Delta x_b \left(\frac{q}{ C_0 r_0^{4 + \frac{1}{n}} }\right)^{\frac{n}{n+1}
    } + P_y^0  + a P_y^0  f(x_b + x_i).\\
\end{split}
\label{eq:P_Nbubbles}
\end{equation}
At this, one must add the capillary pressure drop $a P_{\sigma}\Delta x_b f'(x_b+x_i)$. Summing the contributions of all the $N$ bubbles and neglecting the pressure drop induced by the Newtonian fluid, we obtain the following flow rate equation, that depends not only on the variable $x_b$, but also on the set of constant values $\{x_i\}$:
\begin{equation}
    q(x_b;\{x_i\}) = C_0 r_0^{4+\frac{1}{n}} \!\left[\frac{\Delta P - \gamma(x_b;\{x_i\})}{N\Delta x_b}\right]^{\frac{1}{n} + 1},
    \label{eq:q_Nbubbles}
\end{equation}
with 
\begin{equation*}
\begin{split}
    \gamma(x_b;\{x_i\}) =& N P_y^0 + a\left(P_y^0 F(x_b;\{x_i\}) + P_{\sigma}\Delta x_b F'(x_b;\{x_i\})\right),
    \end{split}
\end{equation*}
and the function 
\begin{equation}
    F(x_b;\{x_i\}) = f(x_b) + \sum_{i=1}^{N-1} f(x_b + x_i).
    \label{eq:F_general}
\end{equation}
The critical pressure $P_y$ needed for the system to flow is then given by the maximum of $\gamma(x_b,\{x_i\})$ in the interval $0<x_b<l$:
\begin{equation}
     P_y(\{x_i\}) = \max_{0<x_b<l}\left[ \gamma(x_b,\{x_i\})\right].
    \label{eq:Py_general_Nbubbles}
\end{equation}
From eq. \eqref{eq:Py_general_Nbubbles} we can see that the value of the critical pressure relies thus not only on the number of bubbles, but also on the specific configuration of the bubbles position, namely on their distances $\{x_i\}$.
Among the possible ensemble of bubbles configurations, the most relevant is the one where the bubble are evenly distributed.
In the diluted limit where $N \Delta x_b$ is very small compared to the tube length, the position of the $N$ bubbles shifted in the first period is uniformly distributed in the interval $(0,l)$.
Our first goal is to compute the probability distribution function of the critical pressure, $\Pi(P_y)$, associated to such ensemble.\\
The second goal is characterize the flow rate. Again the flow of a given tube averaged over a period, $\langle q(\{x_i\} \rangle_T$, depends on its specific bubbles configuration, and thus on its pressure threshold value $P_y = P_y(\{x_i\})$. For $\Delta P \to P_y^+$:
\begin{equation}
    T \propto \int_0^l\! \frac{dx_b}{\left[\Delta P - \gamma(x_b;\{x_i\}) \right]^{1+\frac{1}{n}}} \propto \left(\Delta P - P_y\right)^{-(\frac{1}{n} + \frac{1}{2})}
    \label{eq:dxdt_Nbubbles}
\end{equation}
and thus $\langle q(\{x_i\} \rangle_T \propto (\Delta P - P_y)^{1/n + 1/2}$ if the tube modulation is analytical, or, more generally, $\langle q(\{x_i\} \rangle_T \propto (\Delta P - P_y)^{1/n + 1 - 1/\beta}$.\\
A particular interesting case is the fiber bundle model, usually adopted to study the flow in porous media \cite{Roy2019}. There, a pressure drop $\Delta P$ is applied to a system of many identical tubes. Every tube is filled with a Newtonian liquid, and $N$ bubbles of non-Newtonian fluid are injected in each tube at random. Thus, we are interested in the mean flow rate per tube of such system for a given $\Delta P$, namely $\overline{\langle q \rangle_T}$ (here the overline stays for the average over the bubble configurations).\\
For $\Delta P$ slightly greater than $N P_y^0$, we expect that the flow rate of every tube of the bundle follows the small flow power-law exponent $1/n + 1 - 1/\beta$ if the pressure drop applied is greater than the pressure threshold of that tube, namely $\Delta P > P_y$, or is null if on the contrary $\Delta P \leq P_y$. Instead, we have tubes in the large flow limit, whose flow rate is described by the second case of eq. \eqref{eq:q_cyl}, only if $\Delta P$ is sufficiently greater than $\widetilde P_y = ((3n+1)/(2n+1))N P_y^0$. Since $N P_y^0 < \widetilde P_y < (3/2) N P_y^0$ for all $n>0$, there's always a finite range of values of $\Delta P$ for which all tubes in the bundle presenting non-null flow obey to the small flow regime. Moreover, $P_y \geq NP_y^0$ but is typically much lower than $\widetilde P_y$, because the fluctuations on the value of $P_y$ are smaller than the difference between $NP_y^0$ and $\widetilde P_y$. The effects on the mean flow rate caused by the non-uniformity of the tubes can then be seen only if $\Delta P$ is sufficiently close to $N P_y^0$. In this limit we can compute the mean flow rate per tube as
\begin{equation}
\overline{\langle q \rangle_T} \propto \int_{NP_y^0}^{\Delta P} \!dP_y\, \Pi(P_y)  ( \Delta P-P_y)^{\frac{1}{n} +1-\frac{1}{\beta}}.
\label{eq:q_ensemble_avg}
\end{equation}

\subsection{Sinusoidal geometry}
In this section, we study the case
\begin{equation}
    f(x) = \cos ( 2\pi x/l)
    \label{eq:f_cos}
\end{equation}
It is useful to introduce the angle variables $\theta_b = 2\pi x_b/l$ and  $\theta_i = 2\pi x_i/l$. Using the trigonometric relations, we can write
\begin{equation}
    F(\theta_b;\{\theta_i\}) = \cos(\theta_b) + \sum_{i=1}^{N-1} \cos(\theta_b + \theta_i) = A \cos(\theta_b + \phi)
\label{eq:F_cos}
\end{equation}
where the amplitude is
\begin{equation}
    A = \sqrt{\left(1 + \sum_{i=1}^{N-1} \cos\theta_i \right)^2 + \left(\sum_{i=1}^{N-1} \sin\theta_i\right)^2}
    \label{eq:A_cos}
\end{equation}
and the phase shift $\phi = \arcsin\left( \sum_{i=1}^{N-1} \sin\theta_i/A \right)$.
Similarly, we obtain $F'(\theta_b) = - (2\pi/l)A\sin(\theta_b + \phi)$.
So $\gamma(\theta_b,\{\theta_i\})$ can be written as a cosine function
\begin{equation}
    \gamma(\theta_b;\{\theta_i\}) = NP_y^0 + A\, P_{\gamma} \cos (\theta_b + \phi + \varphi)
    \label{eq:Gamma_cos}
\end{equation}
where $P_{\gamma} = a\sqrt{(P_y^0)^2 + (2 \pi P_{\sigma}\Delta x_b/l)^2}$ and $\varphi = -\arccos\left(P_0^y/P_{\gamma}\right)$, from which it's easy to see that the pressure threshold is 
\begin{equation}
P_y = NP_y^0 + A \,P_{\gamma}
    \label{eq:Py_cos}
\end{equation}
 We now discuss three different possible cases related to different configurations of the bubble positions:
\begin{itemize}
    \item Each bubble is separated from its nearest neighbours by a distance equal to the spatial period $l$. This means that $\theta_i = 0\ \forall\ i$, and consequently $P_{y}$ reaches the highest possible value
    \begin{equation}
    P_{y} = N\left(P_y^0 + P_{\gamma}\right)
    \label{eq:Py_cos_max}
    \end{equation}
    \item Each bubble is separated from its nearest neighbours by half of the spatial period $l/2$, so $\theta_i = \pi$ for $i$ odd and $\theta_i = 2 \pi$ for $i$ even. $P_{y}$ takes the lowest possible value
    \begin{equation}
    P_{y} = \begin{cases}
    \hspace{0.6cm} NP^0_{y} \hspace{0.6cm} \text{if } N \text{ even}\\
    \\
    \hspace{0.2cm} NP^0_{y} + P_{\gamma} \hspace{0.2cm} \text{if } N \text{ odd}
    \end{cases} 
    \label{eq:Py_cos_min}
    \end{equation}
    \item The position of every bubble is uniformly distributed along the tube. This is equivalent to suppose that all the $N-1$ angular differences $\theta_i$ are uniformly distributed in the interval $[0,2\pi]$. In the limit of $N$ sufficiently large, $P_{y}$ follows, in the interval $[NP^0_y,+\infty[$, the probability distribution
    \begin{equation}
    \Pi(P_y) =  \frac{2(P_y - NP^0_y)}{NP_{\gamma}^2}\,e^{ -\frac{\left(P_y - N P^0_y \right)^2}{NP_{\gamma}^2} }.
    \label{eq:Py_cos_pdf}
    \end{equation}
\end{itemize}

\begin{figure}
    \centering
    \includegraphics[scale=0.55]{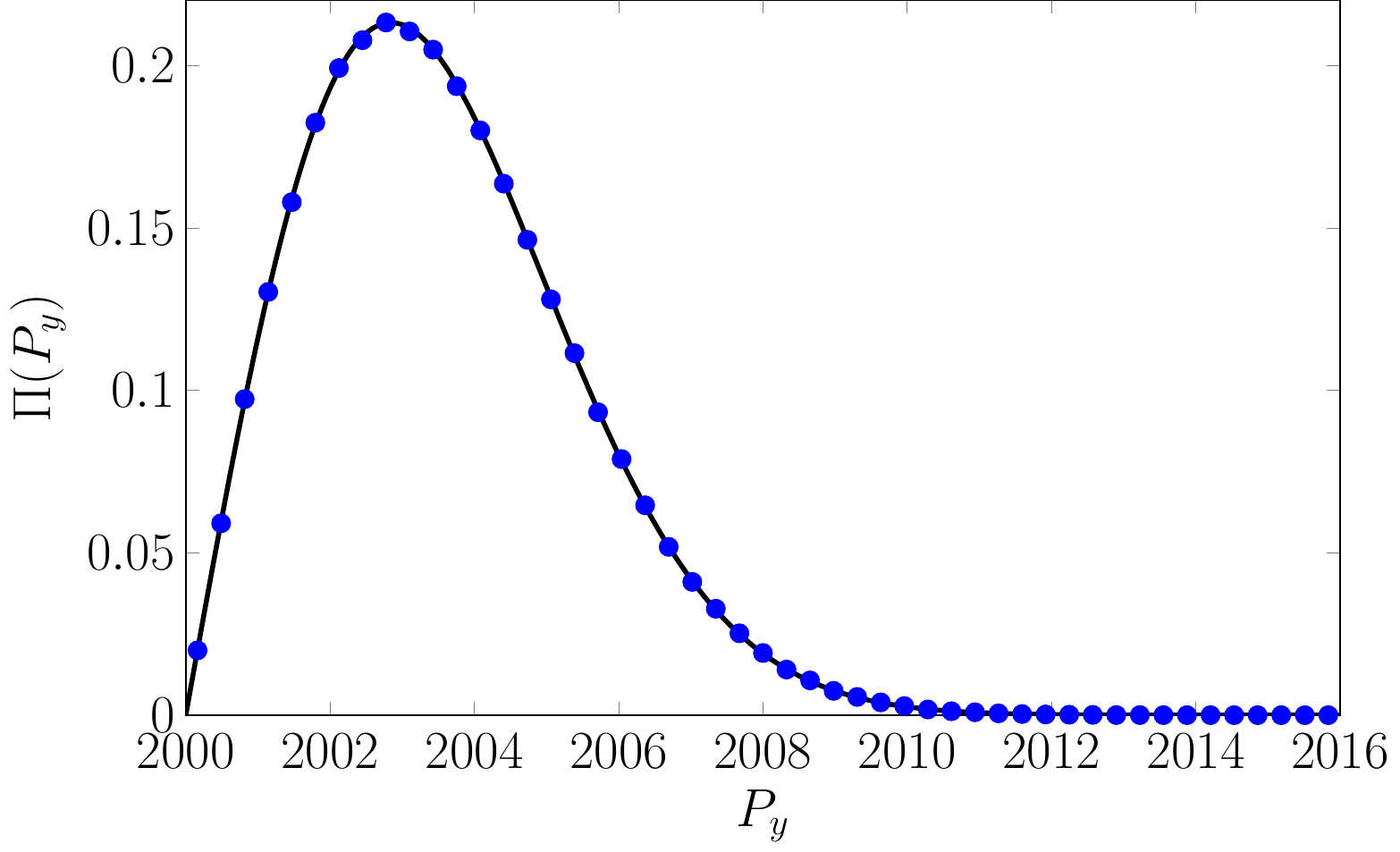}
    \caption{Critical pressure distribution for $N=1000$ bubbles in a sinusoidal tube. Blue dots are the histogram of the numerical data obtained by computing $P_y$ using equations (\ref{eq:A_cos}) and (\ref{eq:Py_cos}) with $\theta_i$ uniformly distributed in $(0,2\pi)$. The solid line is the analytical prediction of equation (\ref{eq:Py_cos_pdf}). Here we set $P_y^0 = P_{\sigma} = 2$, $l=1$ and $a = 0.01$.}
    \label{fig:bubblestube_cos_Py_N1000}
\end{figure}

In order to prove eq. (\ref{eq:Py_cos_pdf}), we first calculate the probability distribution of the variable $B = A^2$:
\begin{equation}
   g(B) = \frac{1}{(2\pi)^{N-1}} \int_0^{2\pi} \! d\theta_1 \, \dots \int_0^{2\pi} \! d\theta_{N-1}\, \delta \left(B - A^2 \right).
   \label{eq:B_pdf_integral}
\end{equation}
To solve (\ref{eq:B_pdf_integral}) it's convenient to perform a Laplace transform:
\begin{equation}
\begin{split}
   \widetilde{g}(s) &= \int_{0}^{+\infty}\! dB\ e^{-sB} g(B)\\
   &= \frac{1}{(2\pi)^{N-1}} \int_0^{2\pi} \! d\theta_1 \, \int_0^{2\pi} \! d\theta_2 \, \dots\\ &\dots \int_0^{2\pi} \! d\theta_{N-1}\, e^{-s \left(\left(1 + \sum_{i=1}^{N-1} \cos\theta_i \right)^2 + \left(\sum_{i=1}^{N-1} \sin\theta_i\right)^2\right)}
\end{split}
\label{eq:gs_integral}
\end{equation}
We define $m_x = \sum_{i=1}^{N-1} \cos\theta_i$ and $m_y = \sum_{i=1}^{N-1} \sin\theta_i$.
Note that the average and the variance of both $\cos\theta_i$ and $\sin\theta_i$ in the interval $[0,2\pi]$, are respectively $0$ and $1/2$.
Moreover, their crossed integral (the covariance) in the same interval is zero, meaning that $m_x$ and $m_y$ are statistical independent. According to the central limit theorem, when $N-1\simeq N$ is sufficiently large, the distribution of both $m_x$ and $m_y$ is Gaussian with mean zero and variance $N/2$.
Eq. \eqref{eq:gs_integral} can be rewritten as
\begin{eqnarray}
   \widetilde{g}(s) &&= \int_{-\infty}^{+\infty} \! dm_x\,\frac{e^{-\frac{m_x^2}{N}}}{\sqrt{\pi N}} \int_{-\infty}^{+\infty}\! dm_y\,  \frac{e^{-\frac{m_y^2}{N}}}{\sqrt{\pi N}} e^{-s \left((1 + m_x)^2 + m_y^2\right) } \nonumber\\
   &&= \frac{e^{-s+\frac{s^2}{1/N + s}}}{1 + Ns} \xrightarrow{N \gg 1} \frac{1}{1 + Ns}.
\end{eqnarray}
The inverse Laplace transform leads to $g(B)=\exp{(-B/N)} /N$, from which eq. \eqref{eq:Py_cos_pdf} easily follows. From \eqref{eq:q_ensemble_avg} and using $\Delta P \to (NP_y^0)^+$ in eq. \eqref{eq:Py_cos_pdf}, we finally obtain the mean flow rate per tube:
\begin{equation}
     \overline{\langle q\rangle_{T}} \propto (\Delta P - NP_y^0)^{\frac{1}{n} + \frac{5}{2}}.\\
    \label{eq:qmeanPy_cos}
\end{equation}

\subsection{Beyond the sinusoidal geometry: the triangular saw tooth shape}

In the previous section we computed explicitly the distribution of critical threshold (see equation \eqref{eq:Py_cos_pdf}) for a tube tube with a sinusoidal deformation and random located identical bubbles. In particular it comes out that the distribution vanishes linearly at $N P_y^0$. How general is this result?\\
We can prove that the result is still robust if the $N$ bubbles have slightly different sizes (see Appendix \ref{sec:appendix_cos_differentsizes}). However, in this section we show that the shape of the distribution is very sensitive to the analytical properties of $f(x)$.  As an important example, we discuss in detail the triangular saw tooth shape introduced in eq. (\ref{eq:f_triangle}), and we first focus on the fully Newtonian case (for which $\tau_c=0$), and then on the non-Newtonian bubbles case but where capillarity effects can be neglected (for which $\sigma=0$).

\subsubsection{Bubbles of Newtonian fluid}

\begin{figure}
    \centering
    \includegraphics[scale=0.54]{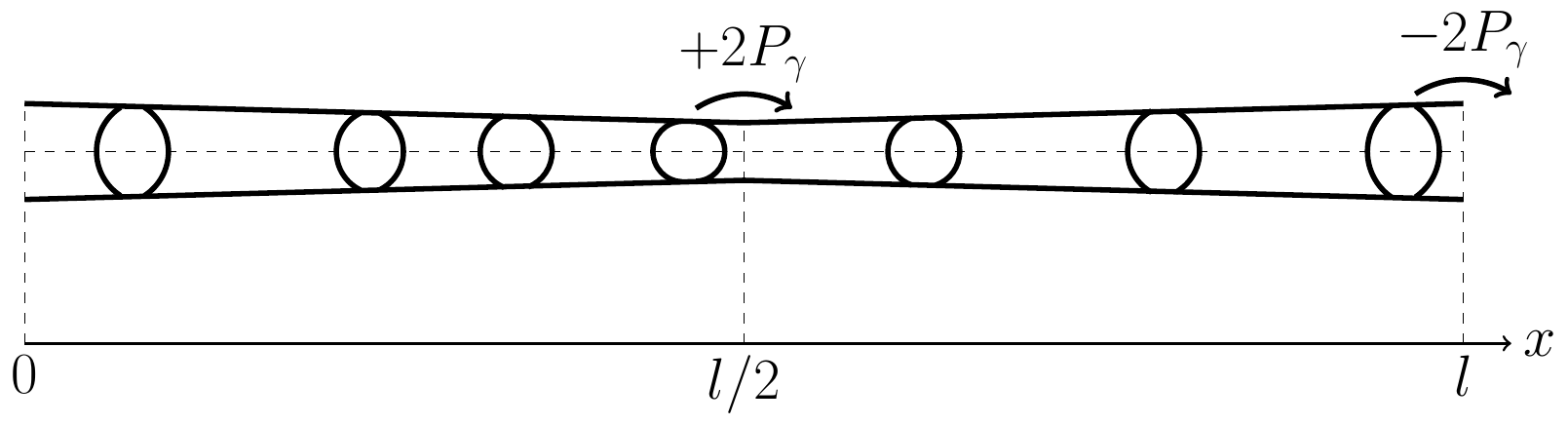}
    \caption{Sketch of several bubbles in a tube presenting the triangular modulation given by eq. \eqref{eq:f_triangle}.}
    \label{fig:sketch_Nbubbles_triangular}
\end{figure}

\begin{figure}
    \centering
    \includegraphics[scale=0.55]{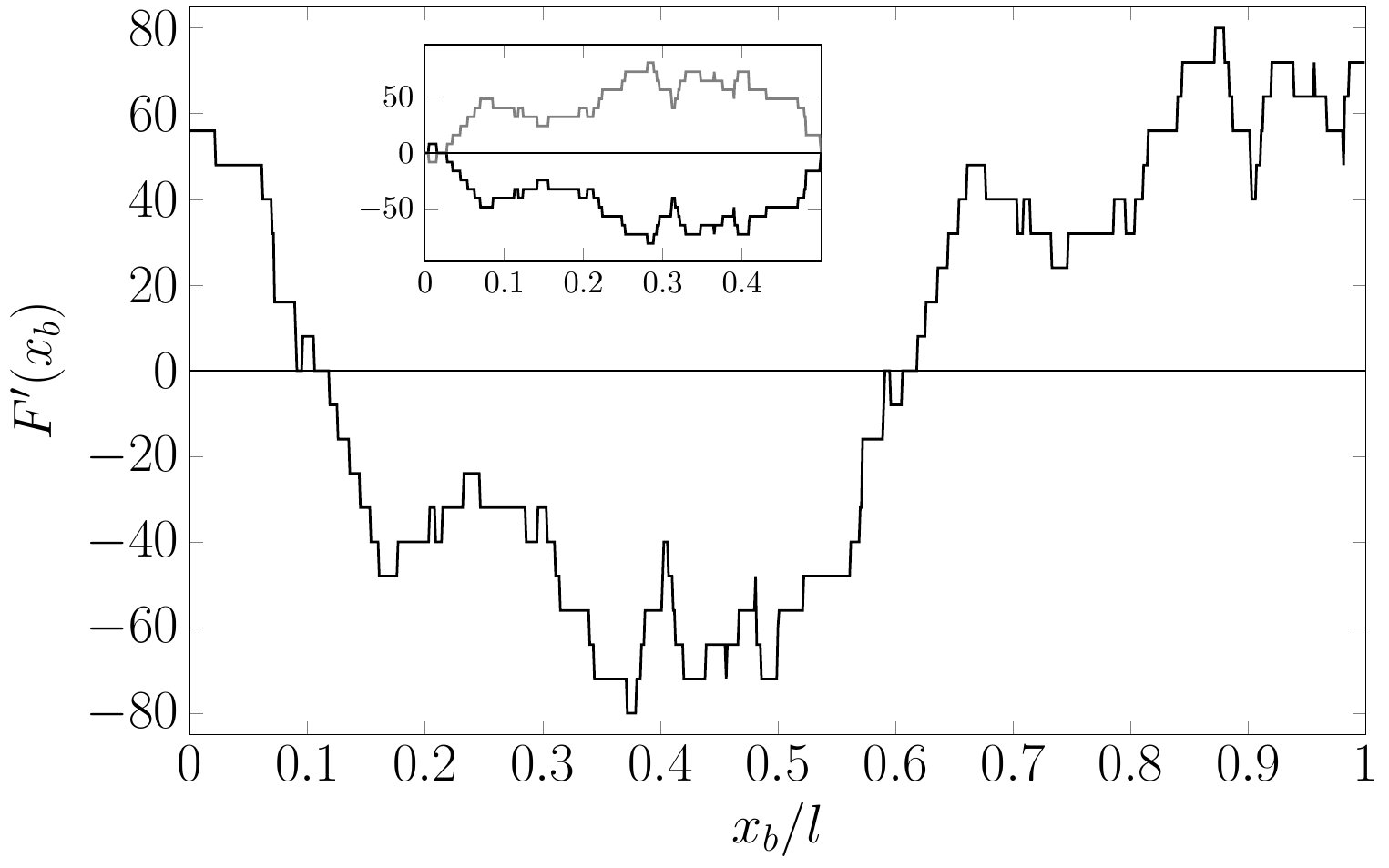}
    \caption{Plot of a typical $F'(x_b)$ in a triangular tube with $N = 50$ bubbles. The two bridges are shown separately in the inset.}
    \label{fig:brownian}
\end{figure}

If the tube is non uniform, even Newtonian bubbles lead to a critical pressure $P_y$, due to the capillary pressure drop at the interface. The value of $P_y$ corresponds to the maximum, in the interval $0<x_b<l$, of the function
\begin{equation*}
    \gamma(x_b;\{x_i\}) = a P_{\sigma} \Delta x_b F'(x_b;\{x_i\}),
\end{equation*}
here $F'(x_b;\{x_i\})$ is the sum of $N$ contributions. For the triangular saw tooth shape, there is a contribution $-4/l$ for every bubble located in the semi-period interval $[0,l/2]$ and $+4/l$ for every bubble in the other semi-period $[l/2,l]$. When $x_b$ moves from $0$ to $l$, all the bubble are shifted of the same quantity. The function $\gamma(x_b;\{x_i\})$ remains constant until
one of the two facts occurs: either the most right bubble belonging to the first semi-period enters the second, so that the function $\gamma$ increases by $2P_{\gamma}$, where $P_{\gamma}= (4/l) a P_{\sigma}\Delta x_b$, or the last bubble belonging to the second semi-period enters the first, so that $\gamma$ decreases by $-2P_{\gamma}$. A sketch of this procedure is shown in figure \ref{fig:sketch_Nbubbles_triangular}.
Increasing $x_b$ further, other jumps occur and the function $\gamma(x_b;\{x_i\})$ performs a periodic random walk in $x_b$ with a diffusion constant $D = 2 P^2_{\gamma}$. A typical trajectory of this random walk is shown in figure \ref{fig:brownian}.
The random walk displays the symmetry $\gamma(x_b;\{x_i\}) =-\gamma(l/2+x_b;\{x_i\})$ and can be decomposed into two Brownian bridges with mirror symmetry, namely two Brownian processes constraint to both start and end at $0$ and with opposite sign. If we denote the two processes $z_1(i)$ and $z_2(i)$, they evolve from $i=0$, in which $z_1(0)=z_2(0)=0$, to $i=N$, in which $z_1(N)=z_2(N)=0$; the two bridges are identical but opposite in sign, namely $z_1(i)=-z_2(i)$. As a consequence, the global maximum of $|\gamma|$ can be written as
\begin{equation}
    P_y = \max_{0<x_b<l/2}\left|\gamma(x_b)\right| = \max_{0<i< N}\left|z_1(i)\right|
    \label{eq:Py_triangF1}
\end{equation}
The exact calculation of the distribution of $P_y$ can be done using the methods discussed in \cite{Mori2020} for Brownian bridges. However, the statistical behaviour of $\max_{0<i\leq N}\left|z_1(i)\right|$ should be similar to the one of the span $S$ of the process, defined as $S =\max_{0<i\leq N}(z_1) -\min_{0<i\leq N} (z_1)$. For the span, rigorous results are proven not only for the Brownian motion but for Gaussian processes with generic Hurst exponent H (the Brownian motion corresponds to $H=1/2$). In particular, the probability to have a small span $\varepsilon$ is known to vanish singularly as \cite{Dean_2014}
\begin{equation}
    \text{Prob}[ S < \varepsilon ] \propto e^{ - k \frac{ D N}{\varepsilon^{1/H}} } \quad \text{for } \; \varepsilon\to 0,
    \label{eq:S_triangle_F1_prob}
\end{equation}
where $k$ is a numerical prefactor of order one. From eq. (\ref{eq:S_triangle_F1_prob}), we can infer that the probability distribution of $P_y$ vanishes as
\begin{equation}
    \Pi(P_y) \propto P_y^{-3} e^{ - \frac{2 k N P^2_{\gamma}}{P_y^{2}} } \quad \text{for } \; P_y\to 0.
    \label{eq:Py_triangle_F1_pdf}
\end{equation}
The presence of an essential singularity at the origin indicates that the tubes with small critical pressure are extremely rare. From eq. \eqref{eq:q_ensemble_avg} we then find that in the limit of small $\Delta P$ the mean flow rate per tube vanishes exponentially as
\begin{equation}
     \overline{\langle q\rangle_{T}} \propto e^{ -\frac{2 k N P_{\gamma}^2}{\Delta P^2} }.\\
    \label{eq:qmeanPy_F1}
\end{equation}

\begin{figure}
    \centering
    \includegraphics[scale=0.55]{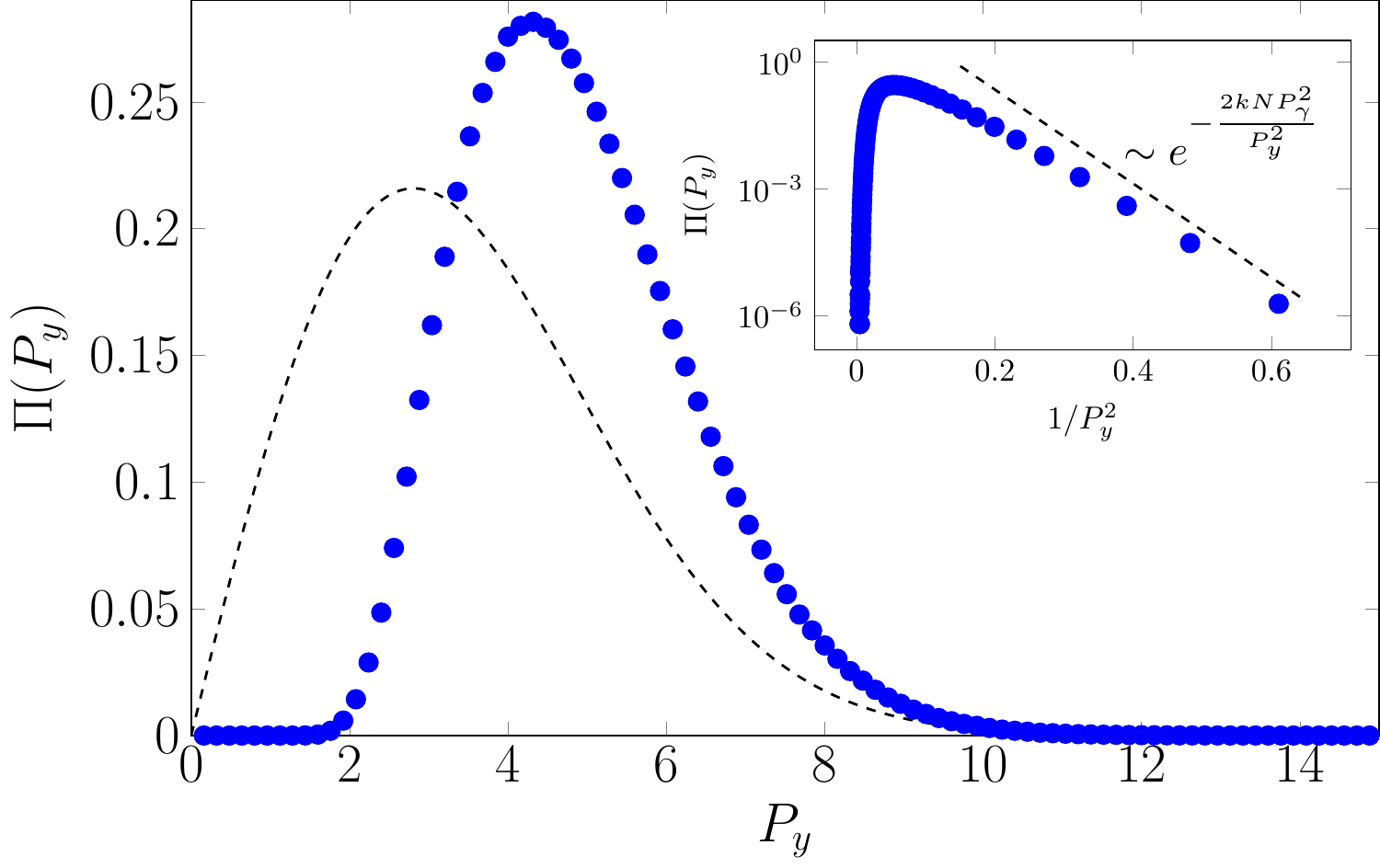}
    \caption{Critical pressure distribution for $N=1000$ bubbles of Newtonian fluid in a triangular tube. Blue dots are the histogram of the numerical data, obtained by generating a Brownian bridge of $N$ steps and extracting its absolute maximum according to eq. (\ref{eq:Py_triangF1}); the dashed curve is the probability distribution (\ref{eq:Py_cos_pdf}) for a sinusoidal tube. In both cases we set $P_y^0 = 0$, $P_\sigma = 2$, $l=1$ and $a = 0.01$. In the inset, the numerical data $(P_y^{-2},\Pi(P_y))$ are compared to the asymptotic trend of eq. (\ref{eq:Py_triangle_F1_pdf}). }
    \label{fig:bubblestube_triangF1_Py_N1000}
\end{figure}

\subsubsection{Bubbles of yield stress fluid without capillary effects}

\begin{figure}
    \centering
    \includegraphics[scale=0.55]{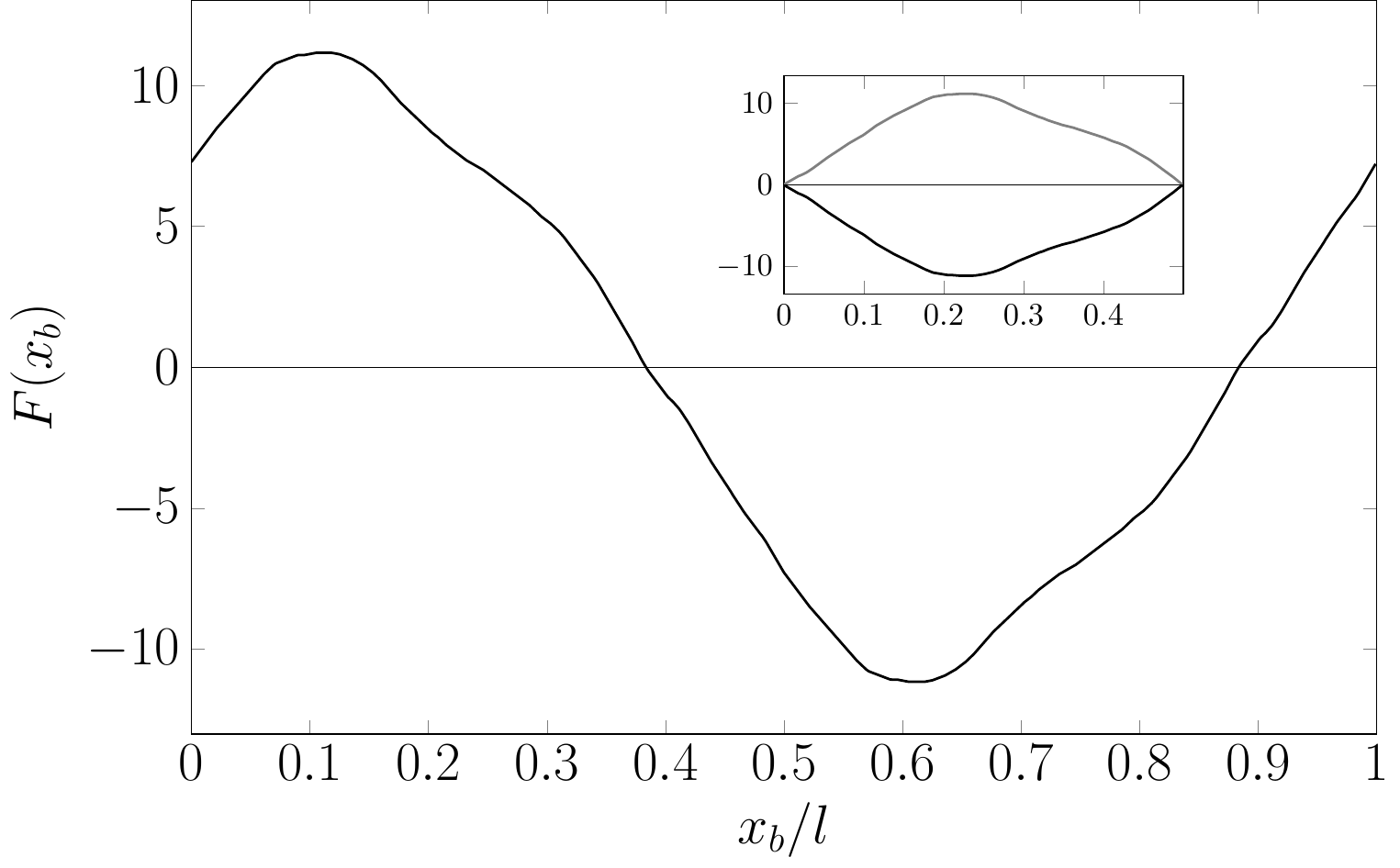}
    \caption{Plot of a typical $F(x_b)$ in a triangular tube with $N = 50$ bubbles. The two bridges are shown separately in the inset.}
    \label{fig:rap}
\end{figure}

The same approach allows to solve the case of bubbles of Non-Newtonian fluid for which we neglect capillary effects. The value of $P_y$ corresponds to the maximum, in the interval $0<x_b<l$, of the function
\begin{equation*}
    \gamma(x_b,\{x_i\}) = NP_y^0 +   a P_{y}^0  F(x_b;\{x_i\}),
\end{equation*}
Here $F(x_b,{x_i})$ is the integral of the random walk discussed in the Newtonian case. A typical trajectory is shown in figure \ref{fig:rap} and corresponds to the trajectory of a Random Acceleration Process (RAP), a piecewise linear function where the slope performs a Random walk; in particular this Gaussian process represents the integral of a Brownian Bridge, and is characterized by $H=3/2$.
The methods discussed in \cite{Majumdar_2010} may be a starting point for deriving an exact form for the distribution of the maximum of a RAP. However, following the lines of the previous discussion, we expect that the distribution of the critical pressure vanishes at $N P_y^0$ as
\begin{equation}
    \Pi(P_y) \propto (P_y - NP_y^0)^{-\frac{5}{3}} e^{ - \frac{k^* NP_{\gamma}^{2/3}}{\left(P_y - NP_y^0\right)^{2/3}} } \hspace{0.3cm} \text{for } \; P_y\to N P_y^0,
    \label{eq:Py_triangle_F_pdf}
\end{equation}
where now $P_{\gamma} = a P_y^0$ and $k^*$ is a numerical prefactor different from $k$. For $\Delta P \gtrsim NP_y^0$ the mean flow rate per tube scales now as
\begin{equation}
     \overline{\langle q\rangle_{T}} \propto e^{ - \frac{k^* NP_{\gamma}^{2/3} }{\left(\Delta P - NP_y^0\right)^{2/3}} }.
    \label{eq:qmeanPy_F}
\end{equation}

As a final remark we note that, as $H \to +\infty$ the function $\gamma(x_b, \{x_i\})$  becomes smoother in $x_b$ and the critical pressure distribution remains singular, but at a higher order of derivative. The linear behaviour in the limit $P_y \to 0$ found for the sinusoidal case represents then the most regular behaviour we can expect.
\begin{figure}
    \centering
    \includegraphics[scale=0.55]{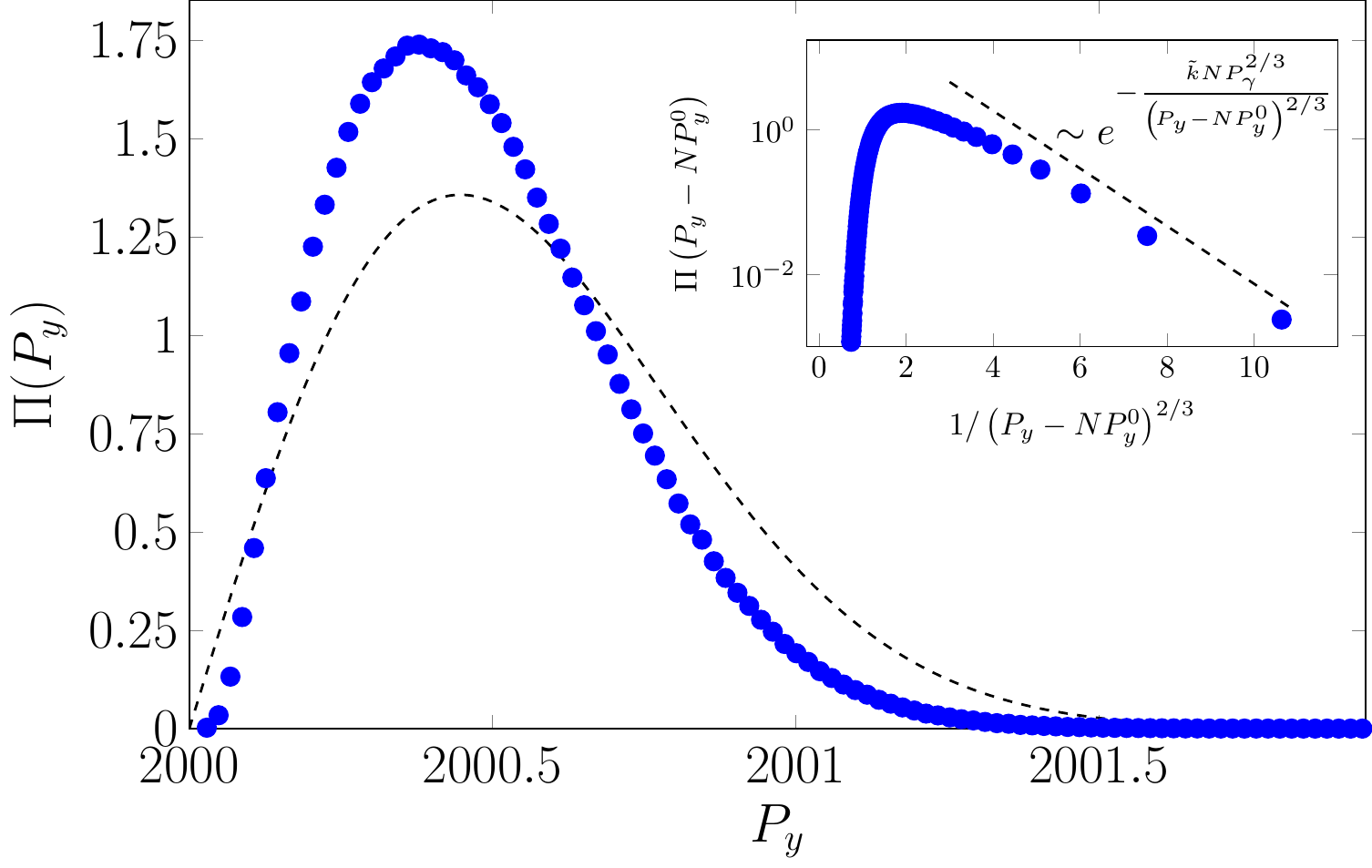}
    \caption{Critical pressure distribution for $N=1000$ bubbles of yield stress fluid in a triangular tube. Blue dots are the histogram of the numerical data, obtained by generating a Random acceleration process of $N$ steps and extracting its absolute maximum; the dashed curve is the probability distribution (\ref{eq:Py_cos_pdf}) for a sinusoidal tube. In both cases we set $P_y^0 = 2$, $P_\sigma = 0$, $l=1$ and $a = 0.01$. In the inset, the numerical data $(\left(P_y - NP_y^0\right)^{-2/3},\Pi(P_y - N P_y^0))$ are compared to the asymptotic trend of eq. (\eqref{eq:Py_triangle_F_pdf}). }
    \label{fig:bubblestube_triangF_Py_N1000}
\end{figure}

\section{Conclusion}

In this paper, we studied the flow rate curve in tubes filled with a Newtonian fluid and where bubbles of non-Newtonian (or Newtonian) fluid are injected.
In presence of a non-Newtonian bubble, or of a non-uniform tube shape, we found a yield pressure threshold, $P_y$, below which there is no flow. Above this threshold, the flow is strongly non-linear and grows with a characteristic exponent:
\begin{equation}
 Q \propto (\Delta P-  P_y )^\beta.
\end{equation}
The value of $P_y$ and $\beta$ depends on the number of bubbles and the geometry of the tubes.
Our results can be summarised as follows:

 \begin{table}[!ht]
            \begin{tabular}{|c|c|c|}
            \hline
                 & Newt& Non-Newt \\
                  \hline
                Uniform & $\beta=1$ & $\beta=1/n + 1$\\
                \hline
              Non-uniform & $\beta=1/2$& $\beta=1/n+1/2$\\
              \hline
            \end{tabular}
            \caption{\label{tab:beta_single} Summary of the exponent $\beta$ when a single bubble is injected in a tube filled with a Newtonian fluid.
            In the case of a non-uniform tube, $\gamma(x)$ is assumed regular around the maximum (\emph{i.e} $\gamma(x)\sim P_y + \alpha (x-x_m)^2$). Otherwise, if $\gamma(x)\sim P_y + \alpha (x-x_m)^\delta$, the exponent are $\beta=1-1/\delta$ and $\beta=1+1/n - 1/\delta$ for the Newtonian and non-Newtonian case respectively.
            }
            \end{table}
            
\paragraph{One bubble:}
            The value of the threshold for the uniform tube of radius $r_0$ is:
            $ P_y= P_{y}^0 = 2\tau_c \Delta x_b /r_0$,
            where $\tau_c$ is the yield stress of the non-Newtonian bubble of size $\Delta x_b$.
            For a non-uniform tube, the value is non-zero even for a Newtonian bubble and given by:
            \begin{equation}
                P_y = P_y^0 +a\max_{0<x_b<l}\left[ P_y^0 f(x_b) + P_{\sigma} \Delta x_b f'(x_b)\right],
            \end{equation}
            where $P_{\sigma}$ is the contribution of the surface tension.
            The values of the exponent $\beta$ in the different configurations are given by table \ref{tab:beta_single}.

    \paragraph{$N$ bubbles:}
    In the case of the uniform tube, the flow curve is identical as for the single bubble. The value of the threshold coincides with the one of a single bubble with the same amount of fluid.
    The case of a non-uniform tube is instead much more interesting. The value of the threshold depends explicitly on the number of bubbles and their relative distance. Assuming that the bubbles are randomly evenly distributed, we derive explicit formulas for the probability distribution of the threshold.
    In particular, for the tube of sinusoidal shape, we found:
        \begin{equation*}
    \Pi(P_y) =  \frac{2(P_y - NP^0_y)}{NP_{\gamma}^2}\,e^{ -\frac{\left(P_y - N P^0_y \right)^2}{NP_{\gamma}^2} }.
    \end{equation*}
    For a fiber bundle model, the total flow curve results from averaging all the bubbles position configurations.
    The values obtained for the exponent $\beta$ are given in table \ref{tab:beta_sinus}.
    
    \begin{table}[!ht]
     \begin{tabular}{|c|c|c|}
             \hline
                   & Newt& Non-Newt \\
                   \hline
                Uniform & $\beta=1$ & $\beta=1/n + 1$ \\
                \hline
               Sinusoidal & $\beta = 5/2$ & $\beta = 1/n + 5/2$  \\
               \hline
            \end{tabular}
            \caption{\label{tab:beta_sinus} Summary of the exponent $\beta$ when many bubbles are injected in a tube with a tube filled with a Newtonian fluid. If the tube is uniform, we recover the result of a single bubble. If the tube has a sinusoidal shape, the exponent is modified.}
            \end{table}
    Note that we consider a sinusoidal deformation; if the tube deformation is less regular, the distribution $\Pi(P_y)$ develops a essential singularity. In the case of the triangular tube we found that the singularity $ \Pi(P_y) \sim \exp{( - 1/P_y^{2} )} $ for the Newtonian bubbles and $\Pi(P_y) \sim \exp{ (- 1/P_y^{2/3}) } $ for the non-Newtonian bubbles in absence of capillary effects.
    
    In conclusion, our study shows that in the case of the fiber bundle model the flow curve can always be described by eq. \eqref{eq:PL}, and we provide an explicit expression for the value of $\beta$ and for the distribution of $P_y$ in different geometry. We remark that within the fiber bundle model the value of $\beta$  depends explicitly on the regularity of the function $\gamma(x)$. One can wonder if this dependence holds also for a realistic porous media.
    Indeed, the fiber bundle model is a crude approximation as all tubes are independent. A challenge for future works is then to solve the flow in frameworks of interacting tubes.

\begin{acknowledgements}
 This work was partly supported by the Research Council of Norway through its Center of Excellence funding scheme, project number 262644. Further support, also from the 
Research Council of Norway, was provided through its INTPART program, project number
309139. This work was also supported by "Investissements d'Avenir du LabEx" PALM (ANR-10-LABX-0039-PALM).
\end{acknowledgements}

\bibliographystyle{plain}

\begin{thebibliography}{10}
\bibitem{note1}
Here we implicitly assume the contact angle between the meniscus and the tube to be small such that the radius of the spherical interface is approximately equal to the radius of the tube.
\end{thebibliography}

\clearpage
\appendix
\section{Bubbles of different sizes in a tube with sinusoidal geometry}
\label{sec:appendix_cos_differentsizes}
We generalize the study of the flow in a tube considering $N$ bubbles of different lengths. We call $\Delta x_0$ the size of the bubble positioned at $x_b$ and $\Delta x_i$ the size of the bubble at $x_b + x_i$, and for all $i$ we take $\Delta x_i \ll l$. We also consider a radius variation small enough so that we can take every $\Delta x_i$ constant. In the limit of small flow rate $q \to 0$, the pressure drop at the edges of the $i$-th bubble is
\begin{equation}
\begin{split}
    &P_{x_b+x_i}^+ - P_{x_b + x_i + \Delta x_i}^- =\\
    &=  \Delta x_i \left(\frac{q(x_b)}{ C_0 r_0^{4 + \frac{1}{n}} }\right)^{\frac{n}{n+1}
    } + P_{y,i}^0\left(1 + a  f(x_b + x_i)\right).\\
\end{split}
\label{eq:P_Nbubbles_differentsize}
\end{equation}
where $P_{y,i}^0 = 2\tau_c\Delta x_i/r_0$. To this, one must add the capillary pressure drop $a P_{\sigma}\Delta x_i f'(x_b+x_i)$. Summing the contributions of all the $N$ bubbles and neglecting the pressure drop induced by the Newtonian fluid, we obtain the following flow rate equation:
\begin{equation}
    \begin{split}
    &\hspace{0.5cm} q(x_b,\{x_i\};\{\Delta x_i\})=\\
    &= C_0 r_0^{4+\frac{1}{n}} \!\left[\frac{\Delta P - \gamma(x_b;\{x_i\},\{\Delta x_i\})}{\sum_{i=0}^{N-1}\Delta x_i}\right]^{\frac{1}{n} + 1},
    \label{eq:q_Nbubbles_differentsizes}
    \end{split}
\end{equation}
where
\begin{eqnarray}
    &&\gamma(x_b;\{x_i\},\{\Delta x_i\}) = \sum_{i=0}^{N-1}P_{y,i}^0 +\nonumber\\
    &&+ a\left( \frac{2\tau_c}{r_0} G(x_b,\{x_i\}) + P_{\sigma} G'(x_b,\{x_i\})\right)
\end{eqnarray}
and the function
\begin{equation}
    G(x_b;\{x_i\},\{\Delta x_i\}) = \Delta x_0 f(x_b) + \sum_{i=1}^{N-1} \Delta x_i f(x_b + x_i).
    \label{eq:G_general_differentsizes}
\end{equation}
We now focus on the case of a tube presenting the sinusoidal modulation given by eq. (\ref{eq:f_cos}). Defining $\theta_b = 2\pi x_b/l$ and $\theta_i = 2\pi x_i/l$, equation (\ref{eq:G_general_differentsizes}) can be written as a single sine function
\begin{equation}
     G(x_b;\{x_i\},\{\Delta x_i\}) = A\sin\left(\theta_0 + \phi \right)
\end{equation}
with the amplitude
\begin{equation*}
     A = \sqrt{\left(\Delta x_0 + \sum_{i=1}^{N-1} \Delta x_i  \cos\theta_i\right)^2 + \left( \sum_{i=1}^{N-1} \Delta x_i \sin\theta_i \right)^2}
\end{equation*}
and the phase shift $\phi = \arcsin\left( A^{-1} \sum_{i=1}^{N-1}\Delta x_i \sin\theta_i \right)$. Similarly, we obtain $G'(x_b;\{x_i\},\{\Delta x_i\}) = - A (2\pi/l) \sin(\theta_b + \phi)$. So $\gamma(\theta_b;\{\theta_i\},\{\Delta x_i\})$ can be written as:
\begin{equation}
    \gamma(\theta_b;\{\theta_i\},\{\Delta x_i\}) =  \sum_{i=0}^{N-1}P_{y,i}^0 + A\, P'_{\gamma} \cos (\theta_b + \phi + \varphi),
    \label{eq:Gamma_cos_differentsizes}
\end{equation}
where $P'_{\gamma} = a\sqrt{ (2\tau_c/r_0)^2 + (2\pi P_{\sigma}/l)^2 }$ and $\varphi = -\arccos\left(2\tau_c/(r_0P'_{\gamma})\right)$.The maximum of eq. \eqref{eq:Gamma_cos_differentsizes} gives the pressure threshold
\begin{equation}
P_y = \sum_{i=0}^{N-1}P^0_{y,i} + A \,P'_{\gamma}.
    \label{eq:Py_cos_differentsizes}
\end{equation}
We now suppose that every bubble size is distributed uniformly between two extreme values $\Delta x_{\rm{m}}$ and $\Delta x_{\rm{M}}$, with $\Delta x_{\rm{m}}<\Delta x_{\rm{M}}\ll l$. Then, for $N$ sufficiently large, $\sum_{i=0}^{N-1} P^0_{y,i} = N \left\langle P^0_{y} \right\rangle$ with $\left\langle P^0_{y} \right\rangle = \tau_c\left(\Delta x_{\rm{M}} + \Delta x_{\rm{m}}\right)/r_0 $. Moreover we assume the angular position $\theta_i$ to be distributed uniformly in the interval $[0,2\pi]$. It follows that the probability distribution $\Pi(P_y)$, in the domain $[N\left\langle P^0_y\right\rangle, +\infty[$, has the following expression:
\begin{equation}
    \Pi(P_y) =  \frac{6(P_y - N\left\langle P^0_y\right\rangle)}{N q P'_{\gamma}}\,e^{ -\frac{3\left(P_y - N \left\langle P^0_y\right\rangle \right)^2}{N q P_{\gamma}^{'2}} };
    \label{eq:Py_cos_pdf_differentsizes}
\end{equation}
here we define $q = \Delta x^2_{\rm{M}} + \Delta x ^2_{\rm{m}} + \Delta x_{\rm{m}}\Delta x_{\rm{M}}$. In particular, $\Pi(P_y)$ vanishes linearly as $P_y \to N\left\langle P^0_y\right\rangle$. To prove (\ref{eq:Py_cos_pdf_differentsizes}), we calculate the probability distribution of the variable $B=A^2$
\begin{equation}
\begin{split}
     g(B) &= \frac{1}{(2\pi)^{N-1}}\frac{1}{(\Delta x_{\rm{M}} - \Delta x_{\rm{m}})^{N}} \int_0^{2\pi} \! d\theta_1\, \dots\! \int_0^{2\pi}\! d\theta_{N-1} \\
     &\times \int_{\Delta x_{\rm{m}}}^{\Delta x_{\rm{M}}}\! d\Delta x_0\ \dots \int_{\Delta x_{\rm{m}}}^{\Delta x_{\rm{M}}}\! d\Delta x_{N-1}\, \delta\! \left(B - A^2 \right).
     \label{eq:B_pdf_integral_differentsizes}
\end{split}
\end{equation}
The Laplace transform of eq. (\ref{eq:B_pdf_integral}) is
\begin{equation}
\begin{split}
   \widetilde{g}(s) &= \frac{1}{(2\pi)^{N-1}}\frac{1}{(\Delta x_{\rm{M}} - \Delta x_{\rm{m}})^{N}} \int_0^{2\pi} \! d\theta_1\, \dots\! \int_0^{2\pi}\! d\theta_{N-1} \\
    &\times \int_{\Delta x_{\rm{m}}}^{\Delta x_{\rm{M}}}\! d\Delta x_0\ \dots \int_{\Delta x_{\rm{m}}}^{\Delta x_{\rm{M}}}\! d\Delta x_{N-1}\\
    & \times e^{-s\left( \left(\Delta x_0 + \sum_{i} \Delta x_i  \cos\theta_i\right)^2 + \left( \sum_{i} \Delta x_i \sin\theta_i \right)^2 \right)}.
    \label{eq:gs_integral_differentsizes}
\end{split}
\end{equation}
We now define the statistical variables $m_x=\sum_{i=1}^{N-1}\Delta x_i\cos\theta_i$ and $m_y=\sum_{i=1}^{N-1}\Delta x_i\sin\theta_i$. The mean and variance of both $\Delta x_i \cos\theta_i$ and $\Delta x_i \sin\theta_i$ in the interval $[0,2\pi]\! \times \![\Delta x_{\rm{m}}, \Delta x_{\rm{M}}]$ are respectively $0$ and $q/6$. $m_x$ and $m_y$ are statistical independent since their covariance is zero. When $N-1\simeq N$ is sufficiently large, the distribution of both $m_x$ and $m_y$ is Gaussian with mean zero and variance $Nq/6$. Eq. \eqref{eq:gs_integral_differentsizes} becomes:
\begin{equation}
\begin{split}
   \widetilde{g}(s) &= \int_{-\infty}^{+\infty} \! dm_x \frac{ e^{-\frac{3m^2_x}{N q}}}{\sqrt{ \frac{\pi N q}{3}}} \int_{-\infty}^{+\infty} \! dm_y \frac{ e^{-\frac{3m^2_y}{N q}}}{\sqrt{\frac{\pi N q}{3}}}\\ &\times \int_{\Delta x_{\rm{m}}}^{\Delta x_{\rm{M}}}\! d\Delta x_0\, \frac{e^{-s\left((\Delta x_0+m_x)^2+m_y^2\right)}}{\Delta x_{\rm{M}} - \Delta x_{\rm{m}}} \\
   &= \frac{1}{\left(1 + \frac{N q}{3}s \right)} \int_{\Delta x_{\rm{M}}}^{\Delta x_{\rm{M}}}\! d\Delta_0\,\frac{e^{-\Delta x_0^2\left(s - \frac{s^2}{\frac{3}{N q} + s}\right)}}{\Delta x_{\rm{M}} - \Delta x_{\rm{m}}}\\
   &\xrightarrow{N \gg 1} \frac{1}{1 + \frac{N q}{3}s}
\end{split}
\end{equation}
The Laplace inversion gives 
$$g(B)=\frac{3}{Nq}e^{-\frac{3B}{Nq}},$$
from which eq. (\ref{eq:Py_cos_pdf_differentsizes}) follows.

\end{document}